\documentclass[mathleft,fleqn,%
]{an}
\usepackage{graphicx}
\usepackage[varg]{txfonts}
\begin{document}

\Pagespan{1}{}
\Yearpublication{2016}%
\Yearsubmission{2016}%
\Month{0}%
\Volume{999}%
\Issue{0}%
\DOI{asna.201400000}%

\title{Surface magnetism of cool stars}

\author{%
O. Kochukhov\inst{1}\fnmsep\thanks{Corresponding author:
        {oleg.kochukhov@physics.uu.se}}
P. Petit\inst{2,3}, 
K.\,G. Strassmeier\inst{4},
T.\,A. Carroll\inst{4},
R. Fares\inst{5},
C.\,P. Folsom\inst{2,3},
S.\,V. Jeffers\inst{6},
H. Korhonen\inst{7},
J.\,D. Monnier\inst{8}
J. Morin\inst{9},
L. Ros\'en\inst{1},
R.\,M. Roettenbacher\inst{8,10},
\and
D. Shulyak\inst{6}
}
\titlerunning{Surface magnetism of cool stars}
\authorrunning{O. Kochukhov et al.}

\institute{
Department of Physics and Astronomy, Uppsala University, SE 75120 Uppsala, Sweden
\and
Universit\'e de Toulouse, UPS-OMP, IRAP, F-31400 Toulouse, France
\and
CNRS, Institut de Recherche en Astrophysique et Plan\'etologie, 14 Avenue Edouard Belin, F-31400 Toulouse, France
\and
Leibniz Institute for Astrophysics Potsdam (AIP), An der Sternwarte 16, 14482 Potsdam, Germany
\and
INAF, Osservatorio Astrofisico di Catania, Via Santa Sofia, 78, I-95123 Catania, Italy
\and
Institut f\"ur Astrophysik, Georg-August-Universit\"at G\"ottingen, Friedrich-Hund Platz, D-37077 G\"ottingen, Germany
\and 
Niels Bohr Institute, University of Copenhagen, Juliane Maries vej 30, DK-2100 Copenhagen, Denmark
\and
Department of Astronomy, University of Michigan, Ann Arbor, Michigan 48109, USA
\and
LUPM, Universit\'e de Montpellier, CNRS, Place Eug\`ene Bataillon, F-34095 Montpellier, France
\and
Department of Astronomy, Stockholm University, SE-106 91 Stockholm, Sweden
}

\received{XXXX}
\accepted{XXXX}
\publonline{XXXX}


\abstract{%
Magnetic fields are essential ingredients of many physical processes in the interiors and envelopes of cool stars. Yet their direct detection and characterisation is notoriously difficult, requiring high-quality observations and advanced analysis techniques. Significant progress has been recently achieved by several types of direct magnetic field studies on the surfaces of cool active stars. In particular, complementary techniques of the field topology mapping with polarisation data and total magnetic flux measurements from intensity spectra have been systematically applied to different classes of active stars leading to interesting and occasionally controversial results. In this paper we summarise the current status of direct magnetic field studies of cool stars, and investigations of surface inhomogeneities caused by the field, based on the material presented at the Cool Stars 19 splinter session.
}

\maketitle

\section{Introduction}

Magnetism is an important, yet incompletely characterized and poorly understood, ingredient of stellar physics. Magnetic fields are playing a key role in stellar evolution, including accretion processes in young stars, angular momentum loss, internal mixing. The fields of cool stars govern dynamic, energetic phenomena on stellar surfaces and significantly influence the stellar environments, including planetary systems. Understanding, for example, the cyclic behavior of cool star magnetic fields is critical for assessing possible impact of the solar variability on the terrestrial climate and exoplanet habitability.

An analysis of the Zeeman effect in spectral lines is the only source of direct information about the strengths and topologies of stellar magnetic fields. During recent years significant progress has been made by magnetic broadening and Zeeman-Doppler imaging (ZDI) studies of cool stars. On the one hand, more physically refined and numerically sophisticated analysis techniques were developed. The number of objects studied with these methods has increased significantly. This allowed establishing the presence of magnetic fields in essentially all classes of cool stars and revealing unexpected trends with stellar parameters. Moreover, long-term monitoring of a handful of sun-like stars yielded first direct observations of magnetic cycles. At the same time, some puzzling discrepancies between results of applications of different diagnostic methods have been identified, suggesting that certain aspects of modern observations are not fully understood or even misinterpreted.

The splinter session ``Surface Magnetism of Cool Stars'' at the Cool Stars 19 conference has provided a comprehensive overview of recent results of the direct studies of magnetic fields in cool stars. A special emphasis was given to the discussion of reliability and consistency of different magnetic indicators and to the comparison of the results obtained by different research groups. In this paper we summarise some of the new results presented at this session. We start with a discussion of limitations of the widely used tomographic field topology reconstruction method (Sect.~\ref{zdi}). Two independent tests of magnetic inversions are presented for the simulated Sun-as-a-star spectropolarimetric observations (Sect.~\ref{sdo} and \ref{vector}), allowing a realistic assessment of the degree of field complexity that can be recovered from modern observational data. We then present results of interferometric imaging of dark star spots on the surfaces of cool active stars (Sect.~\ref{interferometry}) and summarise new findings of the magnetic field studies of solar-type stars (Sect.~\ref{solar}), young cool stars (Sect.~\ref{young}), and low-mass stars (Sect.~\ref{low-mass}). The summary and conclusions are given in Sect.~\ref{conclusions}.

\section{Zeeman Doppler imaging: tests and limitations}
\label{zdi}

Zeeman-Doppler Imaging (ZDI) is a powerful technique to map stellar magnetic fields. It has provided (and continues to provide) a wealth of information on stellar large-scale magnetic surface field distributions in the last decades (see, for a review, Donati \& Landstreet 2009; Fares 2014). 

Spectral lines, if formed in the presence of magnetic fields, are polarised. Studying the polarisation of these lines sheds light on the strength of the magnetic field. Zeeman-Doppler Imaging often uses a set of circular polarisation profiles collected over one or several stellar rotations, and converts these profiles into a magnetic map of the stellar photosphere. This tomographic imaging procedure is ill-posed, a regularisation method, such as Maximum Entropy (Brown et al. 1991; Hussain et al. 2000), Tikhonov regularisation (Piskunov \& Kochukhov 2002) or an iterative regularisation method like the Landweber iteration (Carroll et al. 2012), is needed to get a unique magnetic map. In cool stars, the signature of linear polarisation is very small and hardly detected, it is only very recently that detections of linear polarisation were made (see Sect.~\ref{rosen}). Thus, it is not possible to use linear polarisation for every object, and the mapping consists mainly of inverting circular polarisation profiles. We note here that even for circular polarisation in cool stars, a multi-line technique (e.g. Least-Square Deconvolution or LSD, Donati et al. 1997; Kochukhov et. al. 2010 or multi-line SVD reconstruction, Carroll et al. 2012) should be used in order to detect a polarisation signature; the polarisation level in single lines is usually within the observational noise. 

ZDI suffers from several intrinsic limitations. Small-scale fields on the stellar surface are not resolved, strong fields in dark spots are suppressed by their low surface brightness, signatures of small-scale features can cancel out in some field geometries. In polarised light these features are often missed, especially when using circular polarisation only.
The work presented here aims at testing the consequences of these limitations for reliability of reconstructed large-scale magnetic maps. 

\subsection{Magnetic field of Sun-as-a-star from HMI/SDO magnetograms}
\label{sdo}

\begin{figure*}[!th]
\begin{minipage}[c]{.49\linewidth}
      \includegraphics[scale=0.5]{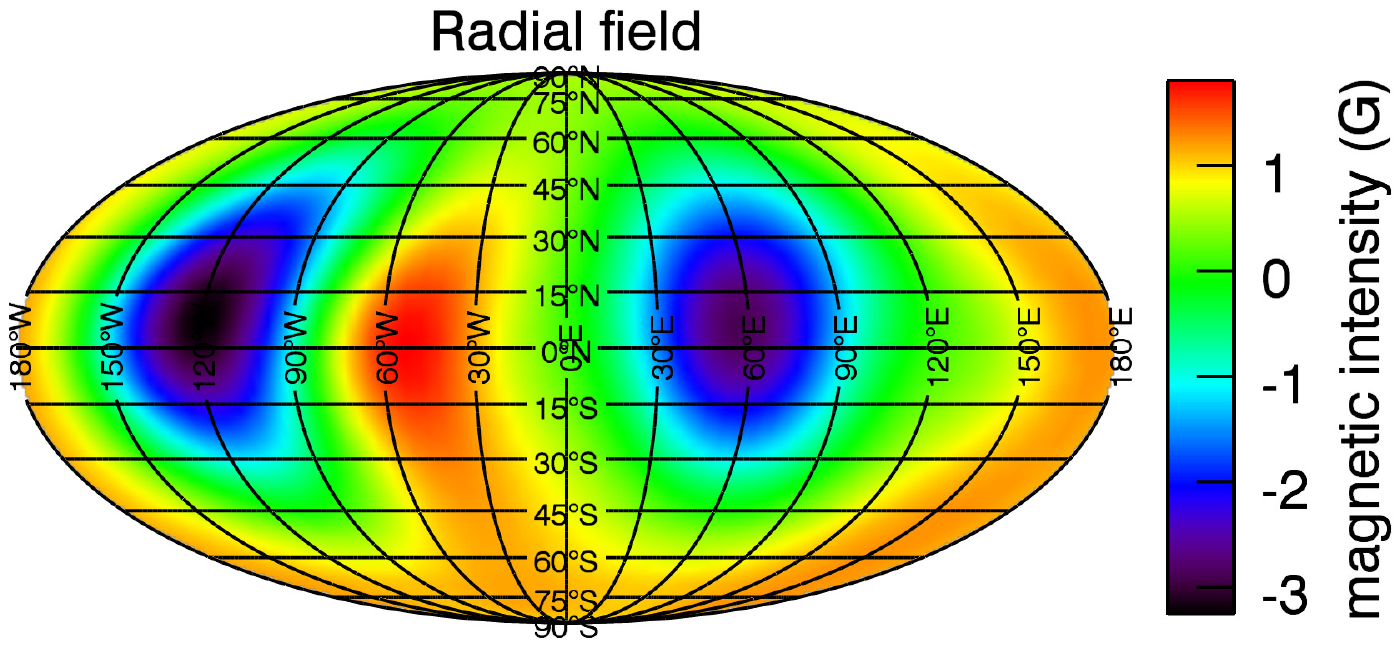}
   \end{minipage} \hfill
   \begin{minipage}[c]{.49\linewidth}
     \includegraphics[scale=0.5]{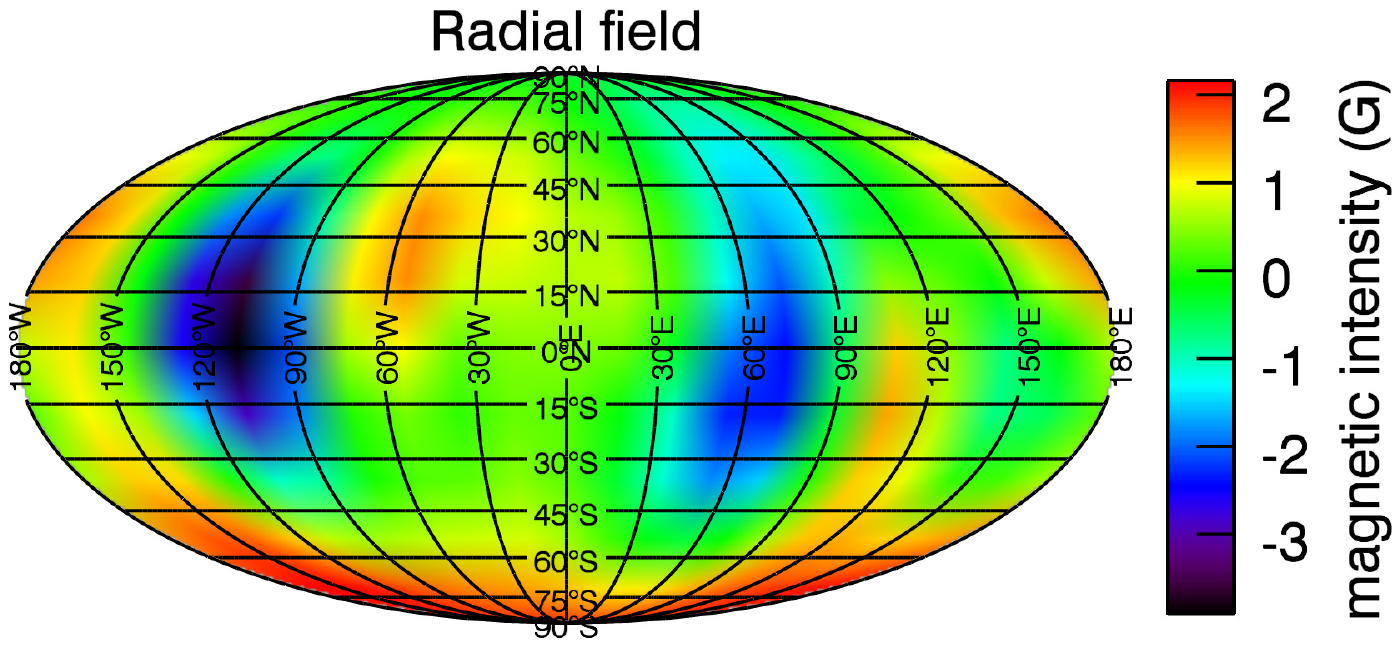}
   \end{minipage}\hfill
\caption{Left: the reconstructed ZDI magnetic map of the Sun, right: the filtered HMI Synoptic map (down to the same spherical harmonics as the ZDI map). The maps show similar configurations (note a difference in the colour scale between the two maps). }
\label{Fig:compar}
\end{figure*}

The Solar Dynamics Observatory (SDO, Pesnell et al. 2012) and the Helioseismic and Magnetic Imager (HMI, Schou et al. 2012) on its board observe the Sun with high cadence. They deliver high-resolution images in which the small-scale field is well resolved. Continuum images, magnetograms and dopplergrams are publicly available. 

Using solar data to test ZDI allows us to assess the effects of small-scale field and its evolution, dark spots, and field geometries on the reconstructed map. Solar HMI/SDO data contain the information we need, both spatial wise and time wise. The Sun, however, is not the best candidate for ZDI. It is a slow rotator, and it is seen nearly equator on. When a star is observed equator on, it is not possible to attribute a feature to the northern or southern hemisphere, which causes a mirroring effect in the reconstructed map. To avoid this affect, we choose an epoch with the highest inclination of the Sun's rotation axis relative to the ecliptic.

To perform ZDI, we developed a technique to produce synthetic intensity and circular polarisation profiles of the Sun-as-a-star, similar to the profiles we collect using stellar spectropolarimeters. A solar intensity profile for one rotation phase is calculated based on the observed intensity and the Doppler maps, taking into account the brightness of each pixel and its velocity. In order to calculate the circular polarisation profile, we use both the calculated intensity profile and the observed magnetogram, and assume a weak-field approximation. To make the synthetic profiles as realistic as possible, we added synthetic noise to the profiles, with signal-to-noise ratios similar to those obtained for solar like stars observed with the ESPaDOnS and NARVAL spectropolarimeters (Marsden et al. 2014).

We then calculated one intensity and circular polarisation profile per day for the Sun over one solar Carrington rotation. We used these profiles as input data for our ZDI code and reconstructed a large-scale magnetic map. The code we used is the ZDI code described in Donati et al. (2006); the magnetic field is represented using a spherical harmonics expansion and Maximum Entropy is used as a regularisation method. 

In order to assess the reliability of the resulting magnetic map, we compared it with the HMI synoptic map for the same Carrington rotation. Synoptic maps are high resolution maps, i.e. having high orders of spherical harmonics. We thus filtered the high order spherical harmonics (small-scale field) from the synoptic map, and kept the low order spherical harmonics (large-scale field) to compare to the large-scale ZDI magnetic map. The ZDI map shows similar features as the filtered synoptic map. In particular, the negative and positive field regions are reconstructed fairly well (see Fig. \ref{Fig:compar}). 

In order to asses quantitatively the similarities between the maps, we calculated, for each bin of longitude, the mean magnetic density by averaging the magnetic field for all latitudes. Our result shows similar magnetic density per longitude bin for the reconstructed map and the synoptic filtered map, with a longitude lag of about $5$~degrees.

The results presented here show that the ZDI radial field is fairly well reconstructed when comparing a ZDI reconstructed map to an observed synoptic map. Small-scale field, adopted from the HMI/SDO data to calculate the circular polarisation profiles fed into the ZDI inversion, does not seem to affect the reconstructed large-scale magnetic field map. This result shows the reliability of the field reconstructed with ZDI. This study will be extended by testing more epochs of observations. The evolution of small-scale and large-scale structures and the effect of such evolution on the reconstructed maps will also be addressed in a future work. 

\subsection{Magnetic field of Sun-as-a-star from vector magnetograms}
\label{vector}

\begin{figure*}
\includegraphics[width=\linewidth]{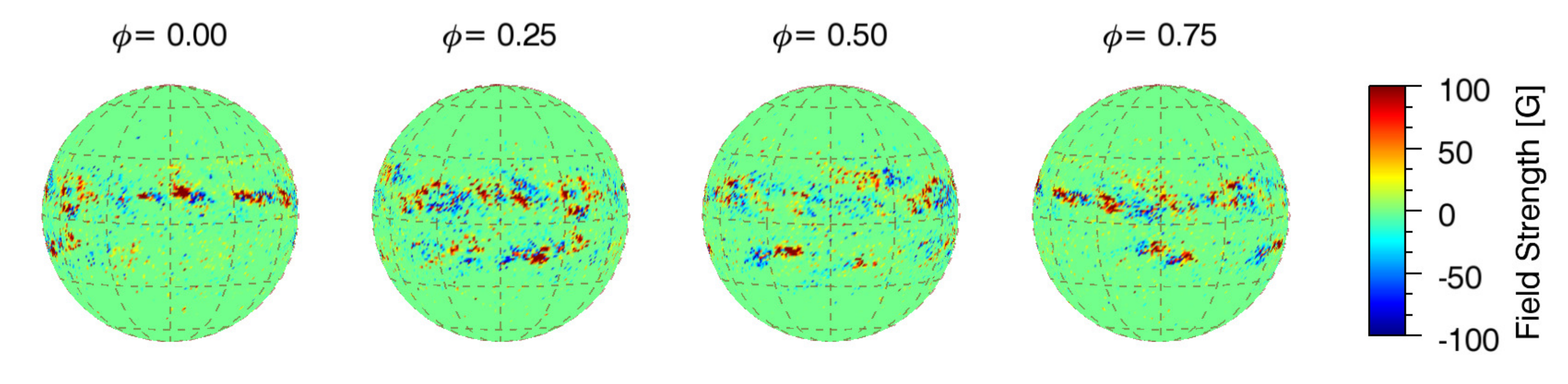}
\caption{Orthographic plot of the high resolution solar synoptic magnetogram (CR 2117). 
The distribution of the radial field is shown for different rotation phases. The surface resolution is 1$^{\circ}$ by 1$^{\circ}$.
The field strength is saturated at 100 G. However, peak values reach up to 1300 G.}
\label{Figure1_Carroll}
\end{figure*} 

\begin{figure*}
\includegraphics[width=\linewidth]{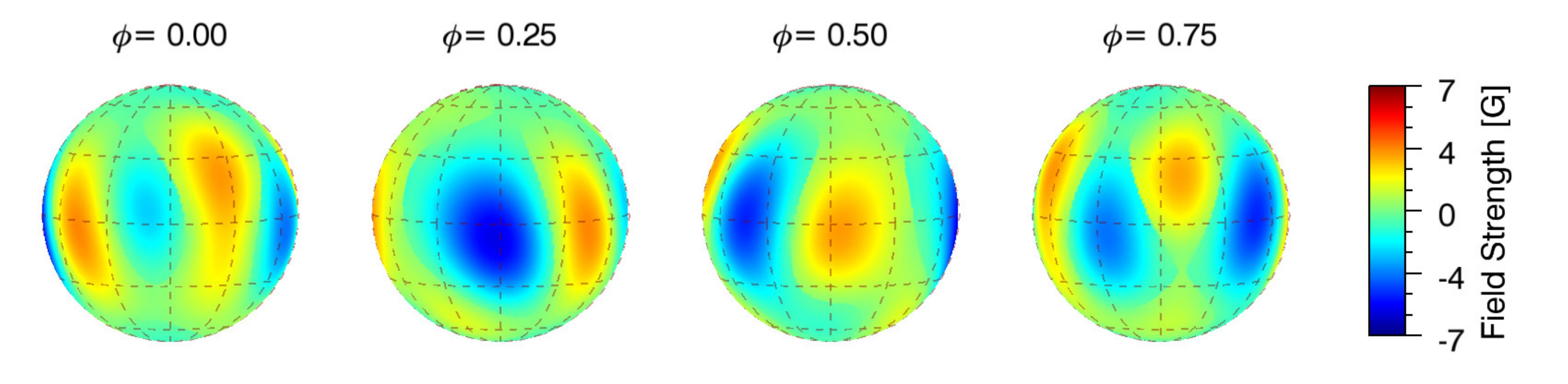}
\caption{Orthographic plot of the reconstructed large-scale solar magnetogram (CR 2117). 
Again the distribution of the radial field is shown for different rotation phases. 
This map is computed by restricting the spherical harmonic reconstruction to $\ell \leq 5$.
The maximum field strength here is 6.8 G.}
\label{Figure2_Carroll}
\end{figure*}

\begin{figure}
\includegraphics[width=\linewidth]{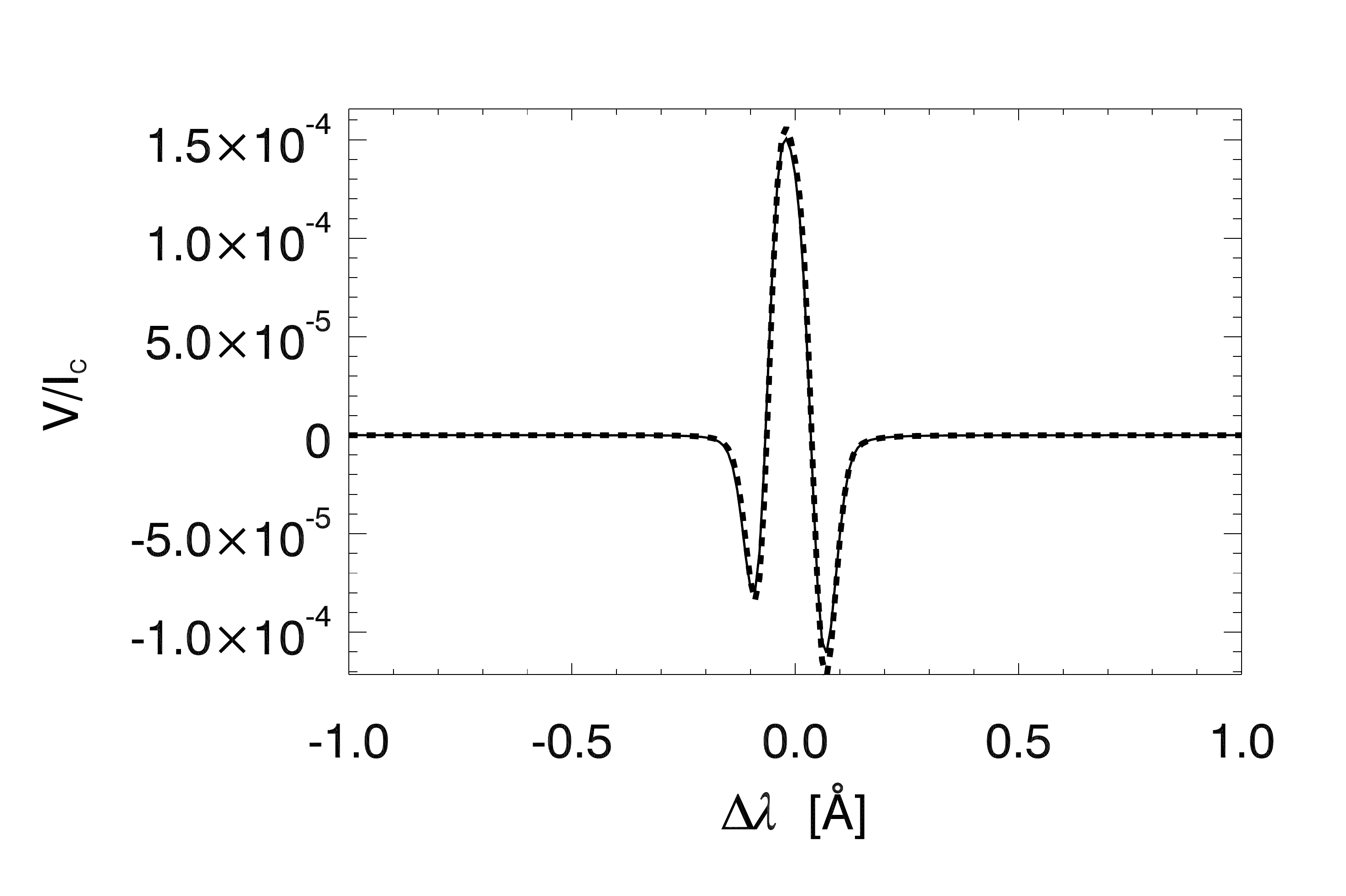}
\caption{Disk-integrated Stokes $V$ profiles resulting from the original map (solid line) 
shown in Fig.\,\ref{Figure1_Carroll},
and from the large-scale, $\ell_{max} = 5$ map (dashed line), from Fig.\,\ref{Figure2_Carroll}. 
The line profiles are computed for the spectral line \ion{Fe}{i} 6173~\AA\ at phase 0.5.}
\label{Figure3_Carroll}
\end{figure}

\begin{figure}
\includegraphics[width=\linewidth]{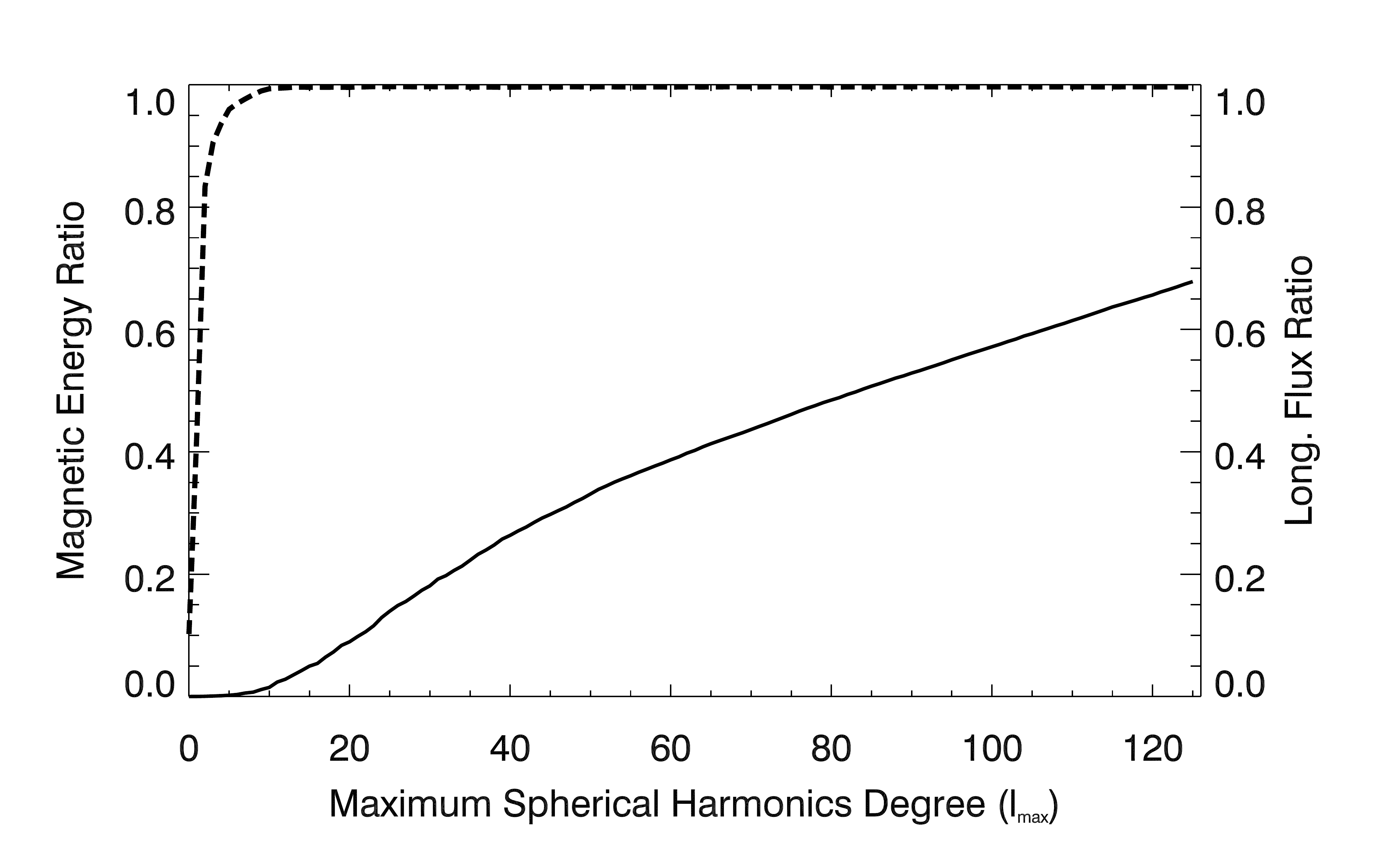}
\caption{Relative magnetic energy content (solid line) and longitudinal magnetic flux (dashed line) over 
maximum degree ($\ell_{max}$). With increasing values of $\ell_{max}$
the energy content of the reconstructed maps slowly rises to the value of the original map. 
The longitudinal magnetic flux on the other hand, which is responsible for the appearance of the 
disk-integrated Stokes $V$ signal, increases much more rapidly to the value of the original map.}
\label{Figure4_Carroll}
\end{figure}

Slow rotating stars can be mapped with ZDI techniques as well as moderate or rapidly rotating stars can. 
However, slow rotation limits the spatial resolution down to a degree that we used to call the large-scale field. 
Although it seems that we all know what is meant when we ascribe a surface field to its
large-scale component, it is not quite apparent what its physical realism is nor its 
relation to the underlying generator (i.e. dynamo) of these fields.

To get an idea about the relation between a detailed magnetic surface distribution and its large-scale 
counterpart we utilised solar SOLIS/VSM synoptic vector magnetograms of Carrington rotation (CR 2117) from Gosain et al. (2013),
to obtain the radial, meridional, and azimuthal component of the vector magnetic field with a surface resolution
of 1$^{\circ}$ by 1$^{\circ}$, see Fig.\,\ref{Figure1_Carroll}. 
This high-resolution magnetogram is then implemented in the ZDI/DI code \emph{iMap} (Carroll et al. 2012) to 
compute the Sun-as-a-star Stokes profiles for various rotation phases.

To obtain the so called large-scale field of the solar magnetogram CR2117, we decomposed the synoptic map 
using a spherical harmonic decomposition. We reconstructed the individual field vector maps by restricting the 
reconstruction to an angular degree of $\ell \leq 5$, see Fig.\,\ref{Figure2_Carroll}.
We used this reconstruction to calculate again the Sun-as-a-star Stokes profiles with the forward module of the
inversion code \emph{iMap}.
The two sets of Stokes $I$ and $V$ profiles calculated from the original high resolution map 
and from the low-order (or large-scale) reconstruction are almost identical for all rotation phases. 
Fig.\,\ref{Figure3_Carroll} demonstrates this for one phase angle. 
In fact, a test ZDI inversion with these profiles -- assuming solar parameters --  yields a ZDI map which has 
striking similarity with the low-order reconstruction, just like the one in Fig.~\ref{Fig:compar}. 
The reason for the similarity between the Stokes $V$ profiles of the original high-resolution map and its 
large-scale reconstruction can readily be understood by realising that the 
longitudinal magnetic flux to the observer is almost the same for both maps, and thus
the disk-integrated Stokes $V$ profiles are the same. One can see this from Fig.\,\ref{Figure4_Carroll}, where
in dashed lines the relative amount of longitudinal magnetic flux generated by the reconstructed maps is shown,
for increasing maximum spherical harmonic degree $\ell_{max}$ of the reconstructed maps.
The solid line represents the relative magnetic energy content as a function of $\ell_{max}$.
These two curves highlight the odd relation between the detailed magnetogram and its large-scale counterpart: 
while both generate the same observable signature already for low $\ell_{max}$ numbers (see the steep increase 
of the dashed curve), they both have a 
vastly different energy content (slow rise of the solid curve).
For $\ell_{max} = 5$ already 95 $\%$ of the longitudinal
magnetic flux (compared to the original high-resolution map) is generated by the large-scale field. 
While there is not even 1$\%$  of the total magnetic energy present in the large-scale reconstruction.

So, did we retrieve the large-scale field of the Sun after all?
What is it, that we see in these low-order reconstructions (or inversions), and how does it relate to the
original high resolution map, or even more complicated to the real underlying field? 
And what kind of information can we extract about the generating process of these fields
if we neglect 99 $\%$ of the magnetic energy?

\subsection{Challenges of cool-star ZDI: self-consistency and four Stokes parameters}
\label{rosen}

\begin{figure*}[!th]
\includegraphics[width=\linewidth]{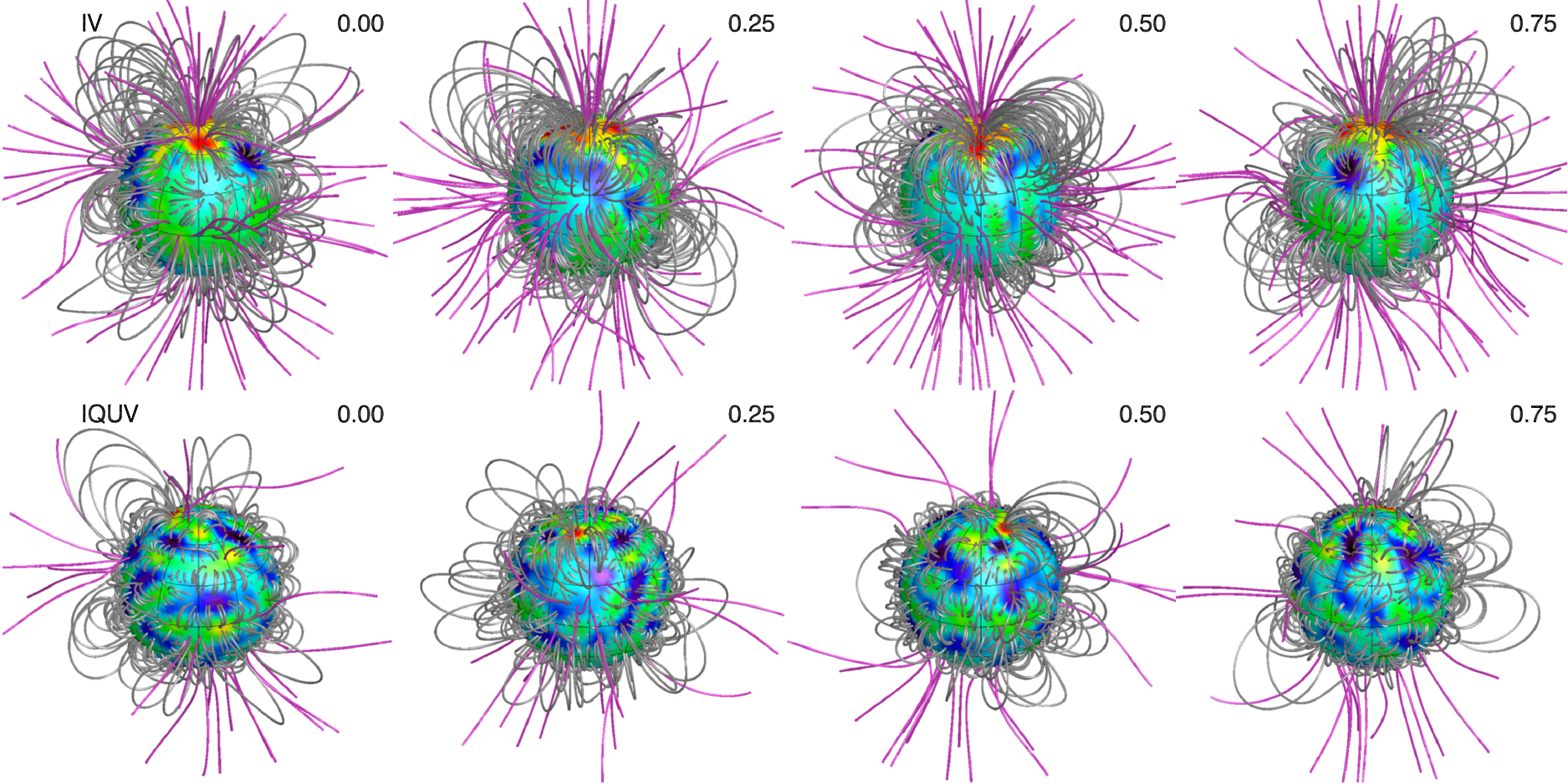}
\caption{Results of the Stokes $IV$ (upper row) and Stokes $IQUV$ (lower row) ZDI inversions for the RS~CVn star II~Peg (Ros\'en et al. 2015). The star is shown at four rotational phases, with the magnetic field lines rendered with the potential field extrapolation using the radial field map reconstructed from observations. The open and closed
magnetic field lines are shown with different colour. The spherical map corresponds to the radial surface magnetic field distribution.
Notice the increase of small-scale fields in the $IQUV$ reconstruction compared to the $IV$ reconstruction.}
\label{fig:rosen}
\end{figure*}

Magnetic fields strongly influence stellar and planetary evolution. It is therefore important to reconstruct the magnetic fields as accurately as possible. The magnetic fields of cool stars are, in general, complex, evolving and relatively weak. Historical cool star magnetic field studies are made by assuming a homogeneous surface temperature, (constant Stokes $I$), and by only using circular polarisation, (Stokes $V$), since that is usually the only type of detectable polarisation signal. However, ZDI studies have shown that using only Stokes $V$ is not optimal (e.g. Donati \& Brown 1997; Kochukhov \& Piskunov 2002; Ros\'en \& Kochukhov 2012). 

If a cool spot coincides with a magnetic feature, which is typical for sunspots, the spot geometry and positions are poorly reconstructed and the magnetic field strength is severely underestimated if only Stokes $V$ is used to map the magnetic field (Ros\'en \& Kochukhov 2012). If temperature inhomogeneities are ignored, the low amplitude of the polarisation profiles, caused by a lower local intensity, can be misinterpreted as a weak magnetic field. Indeed, it was also shown that the quality of the magnetic mapping is improved if Stokes $I$ is included in the reconstruction process in order to simultaneously derive the temperature distribution (Carroll et al. 2012; Ros\'en \& Kochukhov 2012). 

There are also other limitations to Stokes $V$ and Stokes $IV$ mapping. The same set of Stokes $V$ profiles can correspond to different magnetic field configurations since Stokes $V$ is not sensitive to the azimuth angle of the field. It can also lead to crosstalk, especially between radial and meridional field components (Donati \& Brown 1997; Kochukhov \& Piskunov 2002; Ros\'en \& Kochukhov 2012). If Stokes $QU$ parameters are included in the reconstruction, the field strength is increased for all components, especially for the meridional component, and there is almost no crosstalk (Ros\'en \& Kochukhov 2012). 

However, linear polarisation has only been detected in a few cool stars (Kochukhov et al. 2011; Ros\'en et al. 2013) and only for one cool star, II~Peg, has the detected linear polarisation signatures been sufficient for magnetic imaging. Two sets of observations were obtained for II~Peg and these were used to reconstruct the surface temperature and magnetic field of a cool star using Stokes $IQUV$ for the first time (Ros\'en et al. 2015). To enable comparison with usual ZDI studies, reference temperature and magnetic field maps were also derived from the same sets of observations, but only using Stokes $IV$. 

It is preferable to use individual lines instead of LSD profiles when doing ZDI since many individual lines with well known line parameters can be used. Clear distortions due to temperature inhomogeneities could be seen in individual lines in the Stokes $I$ spectra but no clear polarisation signatures were seen in the Stokes $QU$ spectra of II~Peg. In order to derive magnetic field maps, the LSD Stokes $VQU$  profiles had to be used. The common methodology for ZDI using LSD profiles is to apply the single-line approximation, i.e. treat the LSD profile as a single spectral line with some assigned mean line parameters (e.g. Marsden et al. 2011; Kochukhov et al. 2013; Hussain et al. 2016). It has been shown that this approach is appropriate for Stokes $V$ if the magnetic field is weak, $\leq$~2kG, but not for LSD Stokes $QU$ (Kochukhov et al. 2010). A new ZDI methodology described by Kochukhov et al. (2014) was therefore applied. A table of local synthetic LSD profiles corresponding to different magnetic field strengths, orientations, limb angles and temperatures were pre-calculated with the full polarised spectrum synthesis using the same line mask as was used to derive the LSD profiles of the observations. The observed LSD Stokes $VQU$ profiles were then compared directly to the synthetic LSD Stokes $VQU$ profiles meaning no assumptions about the behaviour of the LSD profiles had to be made. The final temperature and magnetic field maps were derived by combining individual-line temperature mapping with LSD profile magnetic field mapping.

The results show discrepancies between the Stokes $IV$ solution and the Stokes $IQUV$ solution for both sets of observations. The fit between the model profiles and observed profiles was equally good for Stokes $IV$, independently if Stokes $QU$ were also included in the inversion. However, the corresponding LSD Stokes $QU$ model profiles from the Stokes $IV$ inversions do not at all agree with the observed LSD Stokes $QU$ profiles. At the same time, the fit to the Stokes $QU$ profiles is good when Stokes $QU$ are modelled. 

The difference is also seen in the resulting magnetic field maps of II~Peg (see Fig.~\ref{fig:rosen}). The magnetic field is 2.1--3.5 times stronger on average when Stokes $QU$ are incorporated in the inversion compared to using only Stokes $IV$. Since a spherical harmonic decomposition of the magnetic field is used, the complexity of the field can also be compared. The magnetic energy contained in $\ell = 1-5$ is 33--36\% when all four Stokes parameters are used compared to 70--84\% when only Stokes $IV$ are used. Even if the total energy is larger in the Stokes $IQUV$ case, the amount of energy in each $\ell$ is larger in the Stokes $IV$ case for $\ell=1-2$ and $\ell=1-3$ for the two observational sets respectively. For one of the II~Peg observing epochs, the energy of the dipole component in the Stokes $IQUV$ case is only 35\% of the energy of the dipole component in the corresponding Stokes $IV$ inversion. The magnetic energy seems to be systematically shifted toward higher $\ell$ when all four Stokes parameters are used compared to only Stokes $IV$. This implies that Stokes $V$ can be fitted equally well by very different magnetic field configurations, and that Stokes $V$ is not sensitive to complex, high-$\ell$ magnetic field structures, in agreement with the discussion in Sect.~\ref{vector}. 

The extended magnetic field topology of II~Peg was also investigated using the potential source surface extrapolation method (Jardine et al. 2002). The results (see Fig.~\ref{fig:rosen}) showed there were more open field lines in the Stokes $IV$ magnetic map and that the magnetic energy at the source surface was also 2.5--5.3 times higher compared to the Stokes $IQUV$ case. This implies that the magnetosphere of II~Peg is more compact in the Stokes $IQUV$ inversion. This finding has important implications for the stellar wind models and angular momentum loss.

\section{Imaging stellar surfaces with near-infrared interferometry}
\label{interferometry}

For decades observations of spots on stars other than the Sun have been obtained through indirect means, both by using photometry and from high-resolution spectra with Doppler imaging techniques (see Strassmeier 2009 for a review). From the beginning of the starspot studies, it has been an aspiration for many to directly image starspots, and until very recently the Sun was the only star on which cool, dynamo-created starspots had been directly imaged. 

Unfortunately, stars appear spatially very small from Earth and are only seen as point sources by our single-aperture telescopes. The largest stars with dynamo-created cool spots have angular sizes of $\theta \sim$2.5 milliarcseconds (mas), which is smaller than even the pixel scale of the highest-resolution imagers on the largest telescopes (e.g., SPHERE at VLT has a pixel scale of 12.25 mas, Beuzit et al. 2008). Therefore, for resolving starspots one must use much larger telescopes or optical/near-infrared interferometry. Wittkowski et al. (2002) investigated the possibility of using VLTI for studying starspots. The first attempt using AMBER at VLT was carried out by Korhonen et al. (2010). Unfortunately, the longest baseline (140m) and the shortest wavelength ($H$-band) currently available at VLTI do not allow for resolving features on active stars.

The breakthrough in directly imaging starspots is only possible with the Georgia State University's Center for High-Angular Resolution Astronomy (CHARA) Array (ten Brummelaar et al. 2005) and the Michigan InfraRed Combiner (MIRC, Monnier et al. 2004). The CHARA Array is an interferometric facility located on Mt. Wilson in California consisting of six 1-meter telescopes and has baselines ranging from 30 to 330 meters.  The CHARA Array can achieve angular resolutions between 0.2 mas ($V$-band) and 0.7 mas ($K$-band).  MIRC is the only instrument at the CHARA Array that can combine light from all the six telescopes, allowing for robust interferometric imaging ($H$-band). MIRC has been very successful in studying distorted fast rotators (e.g., Monnier et al. 2007), interacting binaries (Zhao et al. 2008), the mysterious eclipse of $\epsilon$~Aurigae (Kloppenborg et al. 2010), and components of binary stars up to the flux ratio of 370$\pm$40 (Roettenbacher et al. 2015).

The capabilities of MIRC and the CHARA Array have initiated the giant leap forward in directly imaging starspots. Recently, Roettenbacher et al. (2016) published two interferometric images of the K-giant primary of a RS~CVn binary $\zeta$~Andromedae. With data from 2011 and 2013, both images were obtained from six-telescope CHARA/MIRC observations spanning the star's rotation period ($P_\mathrm{rot} = 17.77$~days). In order to perform image reconstruction on this unique data set, the code SURFING (SURFace imagING; Monnier in prep.) was written.  SURFING makes a global model of the star for an entire rotation -- the almost nightly observations within one observing run are combined into one surface map, analogous to what is done in Doppler imaging. This results in increased surface resolution of 0.025 mas$^2$ per pixel. 

\begin{figure}[!t]
\includegraphics[width=\linewidth]{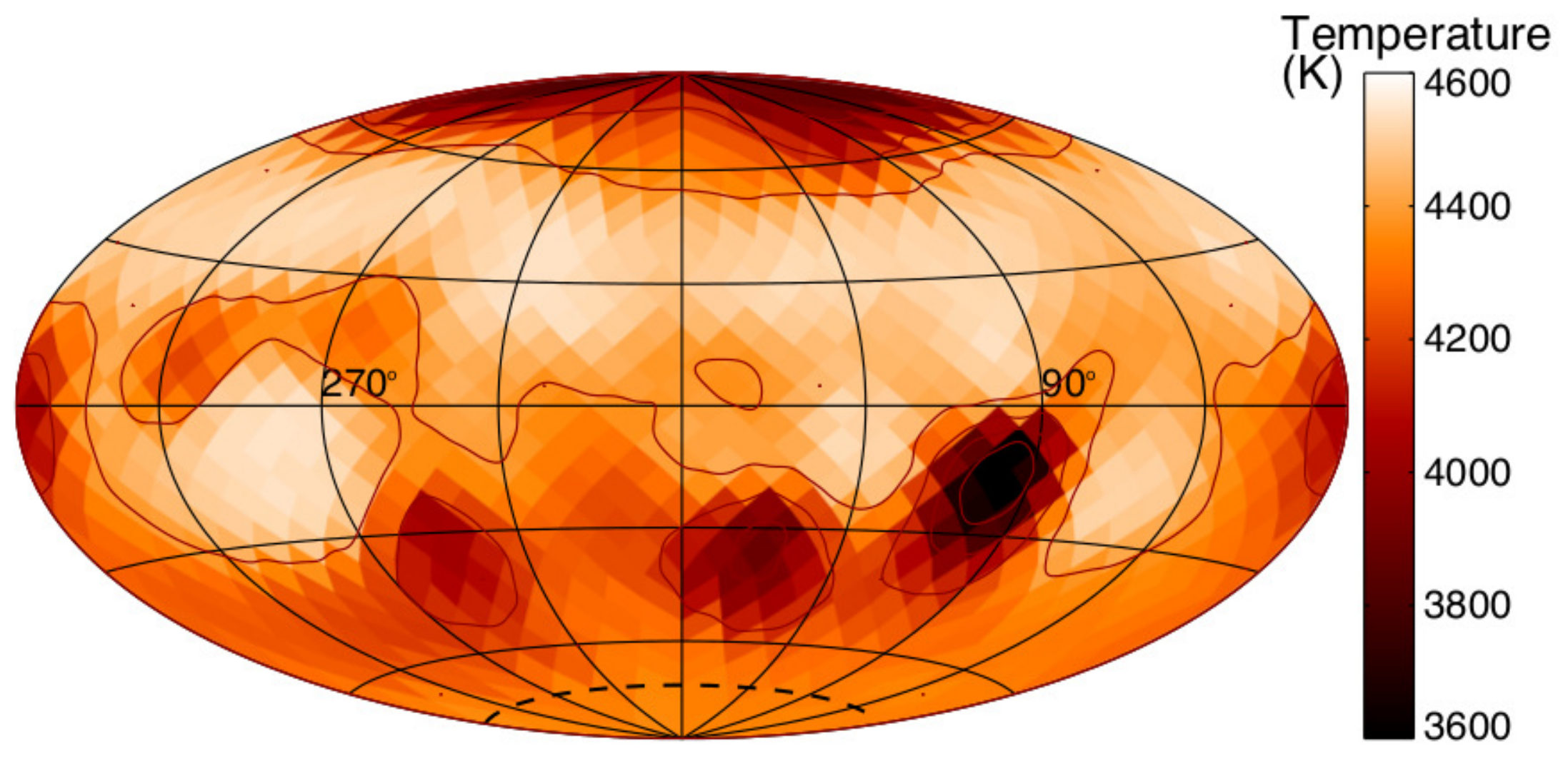}
\caption{
Interferometric image of $\zeta$~And obtained with the SURFING code.  Data were obtained in September 2013 using the MIRC beam combiner at the CHARA Array.  Adapted by permission from Macmillan Publishers Ltd:  Nature. Roettenbacher et al. 2016, Nature, 533, 217, copyright 2016.
}
\label{zetaAnd}
\end{figure}

The interferometric images of $\zeta$~And show clear cool areas on the surface, as shown in Fig.~\ref{zetaAnd}. In both epochs, a prominent polar spot is present, as has also been observed in the previous Doppler images (e.g., Korhonen et al. 2010). Additionally, the maps show lower latitude features that are seen to change significantly between 2011 and 2013. In 2011, the lower latitude cool regions are predominantly located on the Northern hemisphere, and in 2013, they are on the Southern hemisphere. This indicates interesting symmetry breaking in the North-South location of the spots on $\zeta$~And. Only very weak symmetry breaking is seen in the locations of sunspots (e.g., Hathaway 2015). Interestingly, there are indications that during the low solar activity period of 1660--1774, during the so-called Maunder minimum, the solar activity would have shown strong symmetry breaking with spots only on the Southern hemisphere (Ribes \& Nesme-Ribes 1993). Whether this could hint at interesting similarities between the dynamos operating during the grand minima of solar activity and the very active stars like $\zeta$~And is an open topic. In any case, the observed symmetry breaking implies different dynamo operation in $\zeta$~And than in the Sun. However, more data are needed before we can confirm that this symmetry breaking is a persistent phenomenon.

The recent results clearly show that near-infrared interferometry is a new, exciting tool for imaging and studying stellar surface features. It provides the most reliable information on the exact hemisphere on which the spots reside, and can lead to further studies of stellar magnetic structures that were not possible with the previously used indirect methods. At present, interferometric imaging is restricted to only the brightest and closest stars due to limitations in angular resolution, but the method does not require large projected surface rotations, like Doppler imaging does. This opens up new targets that have not been approachable with Doppler imaging.  With present technology, only a small number of spotted targets can be imaged with high enough resolution to reveal discrete spots; however, a larger sample imaged for surface asymmetries as opposed to details can be used to break the hemisphere degeneracies of Doppler imaging results.  

\section{Magnetic cycles of solar-type stars}
\label{solar}

The long-term monitoring of magnetic activity, and in particular the detection of magnetic cycles, in solar-type stars provides an important insight into the mechanisms of dynamo generation and magnetic field amplification. Magnetic cycles in solar-type stars are most commonly investigated using proxies of magnetic activity such as the S-index (Ca II H\&K) used in the Mount Wilson long-term monitoring of chromospheric activity (Wilson 1978; Duncan et al. 1991; Baliunas et al. 1995).  The results of this monitoring show that solar-type stars exhibit different levels of activity variation, irregular activity variations in fast rotating young stars, cyclic activity in comparatively older slowly rotating solar-type stars and Maunder minimum-like flat activity.  

\begin{figure}[!th]
\includegraphics[width=\linewidth]{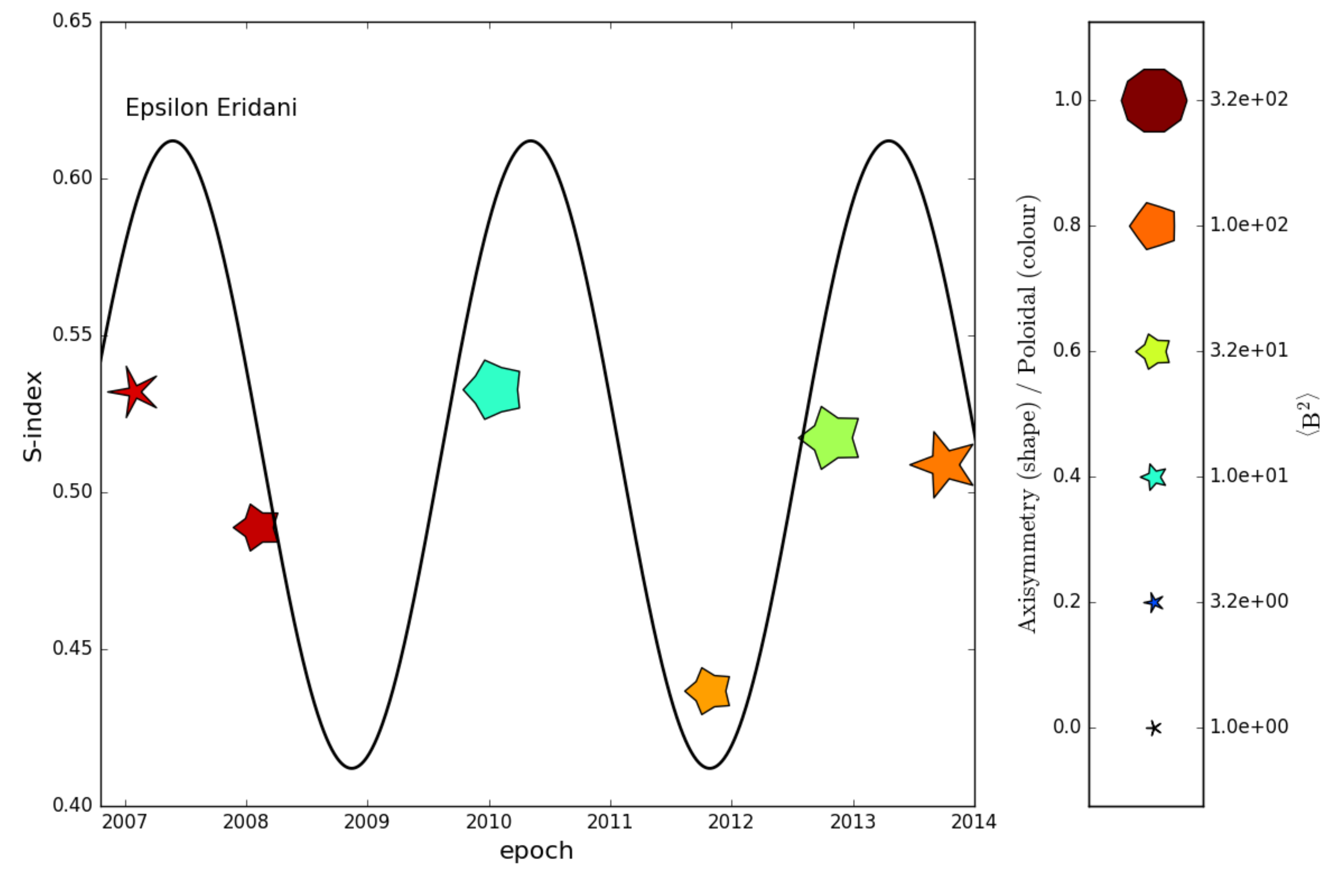}
\caption{The temporal variation of $\epsilon$~Eridani's magnetic field topology as a function of its S-index cycle.   The symbol shape indicates the axisymmetry of the field (non-axisymmetric by pointed star shape and axisymmetric by decagon), the colour of the symbol indicates the proportion of poloidal (red) and toroidal (blue) components of the field and the symbol size indicates the magnetic field strength.}
\label{eps_eri}
\end{figure}

\begin{figure}[!th]
\includegraphics[width=\linewidth]{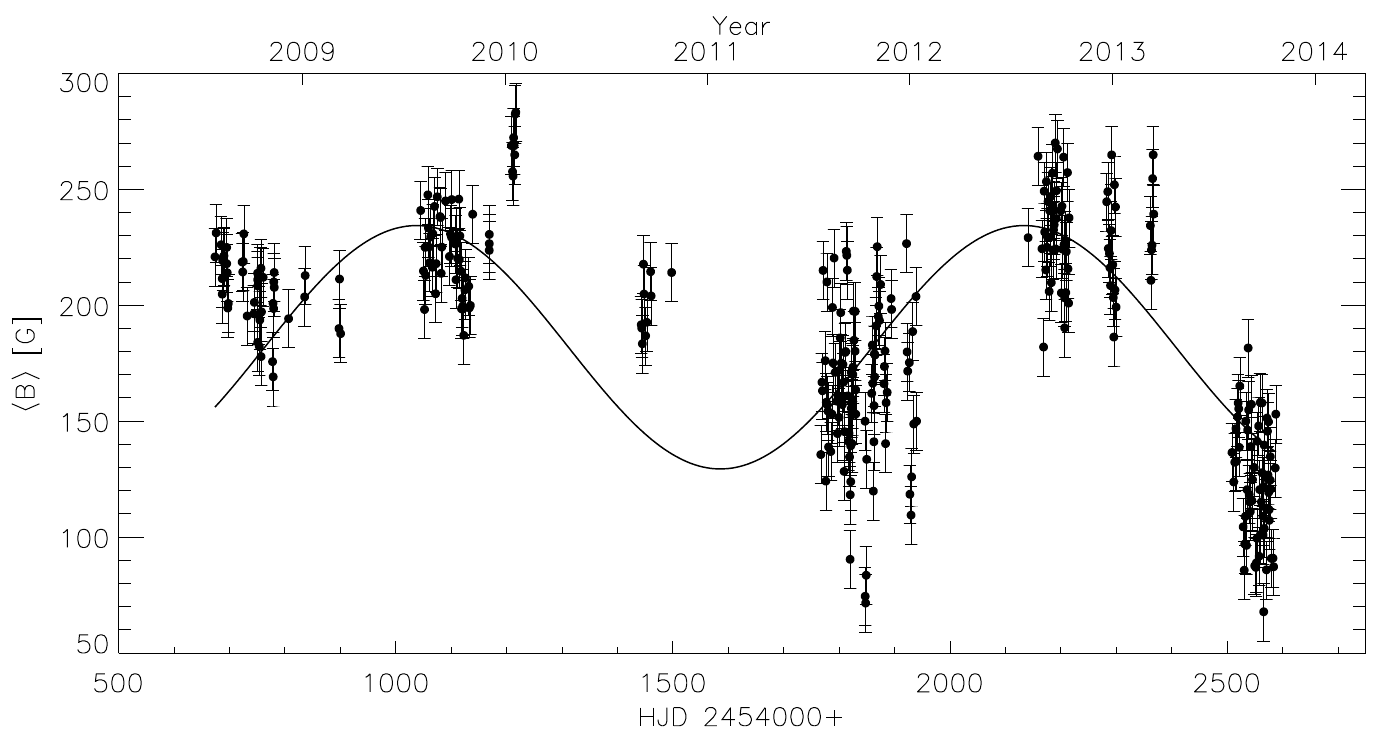}
\caption{Surface averaged magnetic field measurements obtained for $\epsilon$~Eridani by Lehmann et al. (2015) from Zeeman broadening of Stokes $I$ spectra. The solid line indicates a sine-fit using a period of $\approx$\,3~yr.}
\label{eps_eri_kgs}
\end{figure}

While the proxies of magnetic activity are reliable indicators of magnetic activity, they do not give an indication of the geometry of the star's magnetic field.  This can only be obtained using tomographic techniques such as ZDI which uses a time series of spectropolarimetric observations to reconstruct the stellar surface magnetic field geometry.  As the ZDI technique provides information about the
vector magnetic field, it can provide invaluable insights into the large-scale field geometry as well as the temporal evolution and polarity reversals of the large-scale field.

One of the key aims of the BCool collaboration is to monitor the large-scale magnetic geometry over many years. The results show that there are strong variations in the magnetic field geometry and strength of many stars. A short summary of these results is given by Boro Saikia (2016). Here we focus on the evolution of the magnetic field of the K2V planet hosting star $\epsilon$~Eridani ($v\sin i = 2.4$~km\,s$^{-1}$, period\,=\,11.68 days,  age\,=\,2.6 Gyr) and how it varies over its S-index cycle.  $\epsilon$~Eridani's large-scale magnetic field geometry has been reconstructed in Jeffers et al. (2014) over 6 observational epochs (2007--2013).

The results of this long-term monitoring are summarised in Fig.~\ref{eps_eri} where the geometry of the large-scale magnetic field is indicated by the symbol shape and size.   For $\epsilon$~Eridani, the magnetic field geometry is quite variable on timescales of less than a year.  The geometry of the field varies from poloidal to toroidal over similar timescales, though there is no evidence of any cyclic behaviour.  In addition to these results there are additional observations secured in 2014 (at activity minimum) and 2015 (on the approach to activity maximum) and a further analysis of these data will show if there is any possible cyclic behaviour of $\epsilon$~Eridani's large-scale magnetic field.
 
It is instructive to compare results of the spectropolarimetric modelling with the measurements of total magnetic flux from Stokes $I$. 
Such measurements, based on principal component analysis of high and low Land\'e-factor Stokes $I$ line-profiles, were presented for $\epsilon$~Eridani by Lehmann et al. (2015). Clear short-term variations of the surface averaged magnetic field of up to few tens Gauss were detected together with evidence for a three-year cycle (see Fig.~\ref{eps_eri_kgs}). Over time, the grand average surface-field density was $\langle B\rangle = 186\pm47$\,G. The overall trend of these results also fits with the contemporaneous $S$-index measurements from Metcalfe et al. (2013). On the other hand, the field densities reconstructed in the ZDI Stokes $V$ images of Jeffers et al. (2014) were at most $\pm$40\,G for the radial and the azimuthal component and about half of that for the meridional component with a total surface average of at most 20\,G (values ranged between 10$\pm$1 and 20$\pm$1). This is markedly different than the approximately 186\,G from Zeeman broadening analysis and indicates that our respective measuring techniques in Stokes $V$ and Stokes $I$ either suppress or enhance some of the field aspects.


\section{Magnetic fields of young cool stars}
\label{young}

Young solar mass stars undergo a large structural evolution as they traverse the pre-main sequence.  These stars begin their life fully convective, then develop radiative cores as they leave the Hyashi track.  Young stars in this mass range also undergo a significant evolution in their rotation rate.  Early on the pre-main sequence, the stars are strongly interacting with their circumstellar disks, and this interaction regulates their rotation rate.  Eventually the star decouples from its disk, but the star is still on the pre-main sequence and contracting, therefore the rotation rate of the star increases.  These stars also have magnetised stellar winds, and lose angular momentum through the interaction of their magnetic field and wind.  This spin-down is a slow process, thus it only significantly impacts rotation rates after a star has reached the main sequence (e.g. Irwin et al. 2007; Gallet \& Bouvier 2013, 2015).  

The magnetic fields of these stars are expected to be generated by dynamos, most likely through an $\alpha$-$\Omega$ dynamo that depends on both rotation and convection.  Thus both the structural evolution and the rotational evolution should have a strong impact on stellar magnetic fields.  Additionally, stellar spin-down is controlled by the stellar magnetic field, thus an understanding of these magnetic fields is critical for understanding rotational evolution.  

The direct detection of magnetic fields in young fast rotating stars has been achieved through spectropolarimetric observations, detecting the signature of the Zeeman effect in the polarised spectrum of these stars.  Zeeman broadening measurements from total intensity spectra are useful for slower rotating lower mass stars (e.g. Saar 1996; Reiners et al. 2009), but so far these more rapidly rotating stars are challenging targets for that technique.  The spectropolarimetric observations discussed here are from the ESPaDOnS instrument at the Canada-France-Hawaii Telescope in Hawaii, the Narval instrument at the T\'elescope Bernard Lyot in France, and the HARPSpol instrument at the ESO 3.6m telescope in La Silla, Chile.  

A few large surveys of magnetic properties of young solar-like stars have been conducted, or currently are in progress.  The MaPP (Magnetic Protostars and Planets; Donati et al. 2008, 2011) project focused on classical T Tauri stars (cTTS).  The MaTYSSE (Magnetic Topologies of Young Stars and the Survival of massive close-in Exoplanets; Donati et al. 2014, 2015) project is currently in progress, extending this work to weak-liked T Tauri stars (wTTS).  In the framework of the Toupies project (TOwards Understanding the sPIn Evolution of Stars; PI J. Bouvier), Folsom et al. (2016, and in prep.) focused on magnetic fields of older pre-main sequence and young main sequence stars, after most of the structural evolution is complete but spanning the strong rotational evolution of these stars. The Toupies project focused on stars in known open clusters or stellar associations to provide reasonably accurate ages.  The BCool project (Marsden et al. 2014; Petit et al in prep.) focused mostly on older field stars that have spun down significantly, but provides a good comparison for cool stars that are no longer young. These projects all use time series of circularly polarised (Stokes $V$) spectra as input for ZDI to reconstruct the strength and geometry of the large-scale stellar magnetic field.  This methodology has some limitations, most notably small scale magnetic features are below the resolution of the technique and cancel out, leaving them undetected.  However, this is the only method that provides geometric information, and it is the large scale magnetic field that controls the stellar wind and angular momentum loss. 

Using results from these large projects, and some studies of individual stars, Vidotto et al. (2014) found a clear trend of the average large-scale radial magnetic field decreasing with stellar rotation period, as well as Rossby number and age. A number of trends are well established with magnetic activity proxies, such as the X-ray activity Rossby number relation. And a similar trend in Zeeman broadening measurements with Rossby number was found for M-dwarfs (e.g. Reiners et al. 2009).  However, until recently such trends had not been established for large-scale magnetic fields.

\begin{figure}
  \centering
  \includegraphics[width=3.0in]{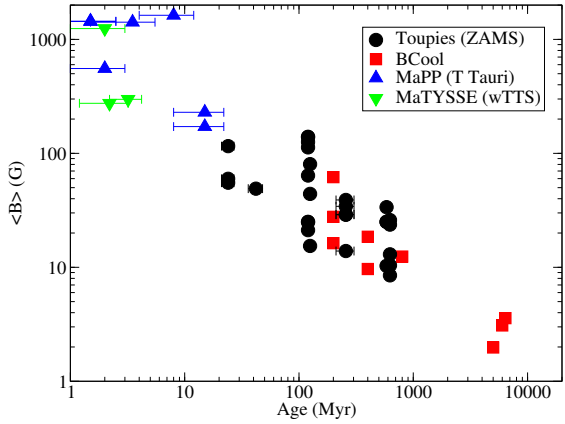}
  \caption{Average large-scale magnetic field strength from ZDI as a function of stellar age.  Data from the Toupies, MaPP, MaTYSSE, and BCool projects are presented here. }
  \label{fig-B-age-correlation}
\end{figure}

\begin{figure}
  \centering
  \includegraphics[width=3.0in]{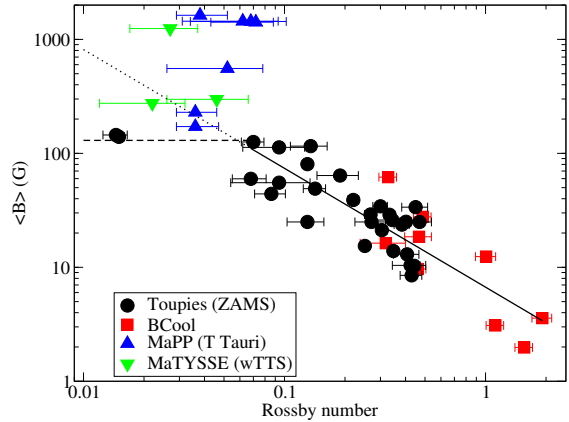}
  \caption{Average large-scale magnetic field strength from ZDI as a function of Rossby number, for the same sample as Fig.~\ref{fig-B-age-correlation}.  The solid line is a power law fit for larger Rossby numbers, and the dashed line is a hypothetical saturation level for small Rossby number.  }
  \label{fig-B-rossby-correlation}
\end{figure}

A similar set of results were found by Folsom et al. (2016), who focused on stars with a narrower range of masses, closer to solar, and with more well determined ages.  They found a continuous decrease in the average (unsigned) magnetic field strength from ZDI with increasing age, approximately following a power law.  This spans T Tauri stars, ZAMS stars, and older main sequence stars, shown in Fig.~\ref{fig-B-age-correlation} (with additional data from Folsom et al. in prep).  Ros\'en et al. (2016) found similar results using a smaller sample of stars, but with multiple epochs of observation for most targets.  For the ZAMS and older main sequence stars, there is a clear power law trend towards decreasing magnetic field strength with increasing rotation period.  The power law relation in rotation period becomes tighter when made in terms of Rossby number (here the ratio of the rotation period to convective turnover time), shown in Fig.~\ref{fig-B-rossby-correlation}.  However, for both period and Rossby number, the two fastest rotators in the sample do not follow the general trend.  They have strengths comparable to more moderately rotating stars, suggesting a saturation of the large-scale magnetic field strength roughly around a Rossby number of 0.1. Vidotto et al. (2014) found some evidence for saturation of magnetic field at low Rossby number based on M-dwarfs, thus due to increasing convective turnover time.  On the other hand, Folsom et al. (2016) present evidence of this saturation due to decreasing rotation period.  

\begin{figure}
  \centering
  \includegraphics[width=3.2in]{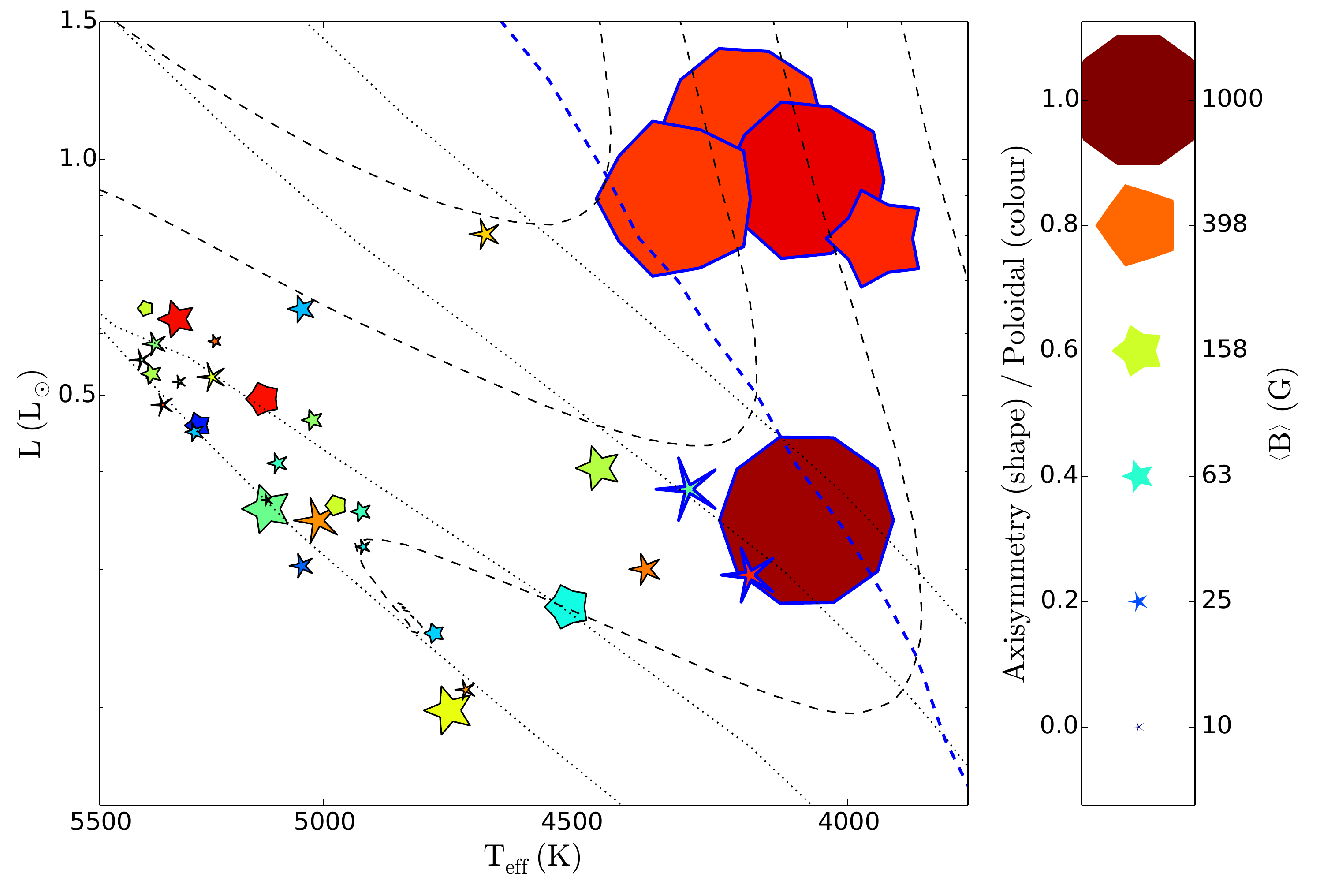}
  \caption{Classical T Tauri stars (thick blue outlines) from the MaPP project and older PMS/ZAMS stars (thin black outlines) from the Toupies project.  Evolutionary tracks (dashed lines) and isochrones (dotted lines) are shown.  The thick dashed blue line indicates where the radiative core has developed to 50\% of the star, by mass.  Symbol size indicates mean magnetic field strength.  Symbol colour and shape indicate how poloidal the magnetic field is and how axisymmetric the poloidal field is, respectively.  }
  \label{fig-hrd-cg}
\end{figure}

The classical T Tauri stars are distinct from the ZAMS stars in the geometry of their magnetic fields, as well as their large magnetic field strengths.  The cTTS magnetic fields are generally mostly symmetric about their rotation axis, dominantly poloidal (as opposed to toroidal), and usually fairly simple.  In contrast, older PMS stars and ZAMS stars have more non-axisymmetric fields, mixes of poloidal and toroidal geometries, and are generally more complex, even for the same Rossby number.  A comparison of cTTSs from MaPP and older PMS and ZAMS stars from Toupies is shown in Fig.~\ref{fig-hrd-cg}.  

The difference in magnetic properties between cTTS and older stars was suggested to be a consequence of their large difference in internal stellar structure by Gregory et al. (2012).  Most of the magnetic cTTSs observed so far are fully convective, or have very small radiative cores.  Thus the convective properties of the stars are likely very different from more evolved solar-mass stars, and they likely do not possess a tachocline.  This is similar to the difference in magnetic properties between fully convective M-dwarfs and partly convective main sequence K stars (e.g. Morin et al 2010).  Further observations of more evolved pre-main sequence stars support this interpretation (Folsom et al. 2016).

Classical T Tauri stars differ from ZAMS stars, not just in their structure, but also in that they are accreting and likely strongly interacting with their disks.  An investigation of (non-accreting) wTTSs in the same parameter range can test whether this impacts surface magnetic fields.  The MaTYSSE project is studying wTTSs, but so far results for only a few stars have been published (e.g. Donati et al. 2014, 2015).  These early results are somewhat inconclusive, in that the wTTS magnetic fields are, on average somewhat weaker and more complex than the cTTS fields.  But they are still stronger for a given Rossby number, and more poloidal and axisymmetric, than the fields of ZAMS stars.  Thus more work needs to be done to extend this sample.

There is now a well established trend of decreasing large-scale magnetic field strength with age, from the PMS through the main sequence.  On the PMS, this trend seems to be largely driven by structural changes in the stars.  On the main sequence, this is largely driven by the rotational evolution of the stars.  There is a good correlation between large-scale magnetic field strength and Rossby number, down to a Rossby number of $\sim$0.1.  In the partly convective stars, there is tentative evidence for saturation of the large-scale magnetic field below a Rossby number $\sim$0.1.  The largely convective T Tauri stars display a distinct set of magnetic properties from the more evolved stars, and this is most likely due to structural differences in the stars. 


\section{Magnetic field of low-mass stars}
\label{low-mass}

Low-mass stars -- understood here as M dwarfs -- have attracted a lot of interest during the past 
few years. In particular, their magnetic fields and activity are at the core of several important 
topics of research. A fundamental issue about M dwarfs is to understand how the dynamo mechanisms 
change from the most massive M dwarfs -- which are partly convective like the Sun -- to the least 
massive ones -- which are fully convective -- and how this change affects the surface magnetic 
field and activity (see e.g. Morin 2012). A better understanding of the magnetic fields of 
M dwarfs is also expected to induce progress on several puzzles such as their rotational evolution 
(e.g. Newton et al. 2016) or the relation between chromospheric and coronal emissions taking place 
at various wavelengths (e.g. Williams et al. 2014). Moreover, M dwarfs have recently become the main 
targets of planet search programs, and the need to better understand and model their magnetism is 
twofold. First, their time-dependent magnetic activity generates radial velocity fluctuations and 
brightness variations which can impede the detection of orbiting planets or even mimic the presence 
of such planets (e.g. Bonfils et al. 2007). Second, knowing the stellar magnetic field, particle wind 
and levels of high energy emission (UV and X-rays), as well as their evolution with stellar age is 
key to assessing the potential habitability of detected planets (e.g. Ribas et al. 2016).

The surface magnetic fields of M dwarfs can be studied using different, and often complementary, 
approaches (see e.g. Morin et al. 2013). More details on the techniques and on results for cool stars 
in general can be found in Reiners (2012) and Morin et al. (2016) for instance. Activity measurements 
correspond to features distributed across the electromagnetic spectrum which generally result from 
the interaction of the magnetic field with the stellar atmosphere. An important result of activity 
measurements of M dwarfs -- either 
chromospheric H$\alpha$ emission, or coronal emission at radio and X-ray wavelengths -- is that 
fully convective mid-M dwarfs follow a rotation-activity relation very similar to that of more 
massive partly-convective stars : the activity level increases toward faster rotation until a 
saturation plateau is reached, for a Rossby number of about 0.1 (e.g. McLean et al. 2012; Wright \& Drake 2016). 
In addition, M dwarfs display long-term variability of their activity levels, with even hints of 
possible activity cycles (Gomes da Silva et~al. 2012).

An alternative and complementary approach consists in directly measuring the magnetic field 
at photospheric level through the Zeeman effect on spectral lines. Such measurements can be carried 
out either in unpolarised or in polarised light, in both cases using high resolution spectroscopy. 
From unpolarised spectroscopy it is possible to derive the average magnetic field of the star. This 
quantity is sometimes referred to as a ``magnetic flux'' -- although it has the dimension of a 
magnetic flux density -- to stress that it does not correspond to the measurement of a \emph{local} 
magnetic field strength on a point of the stellar surface. These measurements are efficient to 
measure magnetic fields regardless of their complexity and have been used to study energetic aspects 
of stellar dynamos (Christensen et al. 2009), but provide very little constraint on the field 
geometry. Using high-resolution and high-signal to noise spectra this method can also provide constraints on the 
distribution of local field strength on the stellar surface (e.g. Shulyak et al. 2014).
This method has been successfully applied to M dwarfs, first using atomic lines for early-
to mid M-dwarfs (e.g. Johns-Krull \& Valenti 1996). With its extension to the FeH molecule, it has 
become possible to measure magnetic fields of stars spanning the whole M spectral type, with low to 
moderate projected rotational velocities (e.g. Reiners \& Basri 2007). These measurements have shown 
that both partly- and fully-convective early- to mid-M dwarfs follow a similar rotation-magnetic 
field relation, exactly as was found for activity measurements. At high Rossby number (slow 
rotation) the measured average magnetic field is anti-correlated with $Ro$, whereas below a saturation threshold 
of $Ro\simeq0.1$ a plateau is observed with magnetic fields of 2--4~kG.

Measuring the properties of the Zeeman effect in polarised light brings again different information 
on stellar magnetic fields. Due to the mutual cancellation of polarised signals arising from 
neighbouring areas of opposite polarities, spectropolarimetry is only sensitive to the large-scale 
component of the field. Conversely to unpolarised spectroscopy, this approach is also sensitive to the 
orientation of the magnetic field vector. The ZDI technique has been 
introduced to make full use of the information contained in spectropolarimetric data (in most cases 
restricted to circular polarisation only, see Sect.~\ref{rosen}): from a 
time-series of polarised spectra sampling at least a stellar rotation cycle, it reconstructs the 
magnetic field vector at the surface of the star (see Semel 1989; Donati et al. 2006). This 
approach has been applied to a sample of active early- to mid-M dwarfs, showing for the first time 
a change in the magnetic properties at the fully-convective boundary (see Donati et al. 2008; Morin et al. 2008). 
Fully-convective stars indeed appeared to generate magnetic fields with a strong 
large-scale component dominated by the axial dipole, while for partly-convective ones the 
large-scale component of the field (i.e. the component probed by spectropolarimetry) is weaker and 
more complex. Interestingly, the latest development of dynamo numerical simulations now reconcile 
the measurements in unpolarised and polarised light, with magnetic fields exhibiting both an 
energetically dominant small-scale component and a large-scale component dominated by the axial 
dipole mode (Yadav et al. 2015). On the observational side, the latest studies are now overcoming 
the limitations of the first samples, with measurements in unpolarised light being extended to 
rapid rotators (see Sect.~\ref{shulyak}) and spectropolarimetric surveys exploring 
moderate rotators in the unsaturated regime (H{\'e}brard et al. 2016).

Going towards later spectral types, very-low mass stars and ultracool dwarfs exhibit 
intriguing behaviours when observed with these different approaches. Their chromospheric activity 
decreases towards late spectral types (e.g. Reiners \& Basri 2010). X-ray and radio luminosities 
exhibit a large scatter for rapid rotators, resulting in a break of the ``G\"udel-Benz'' 
correlation with rapidly-rotating ultracool dwarfs often appearing radio-bright and X-ray faint, 
but the opposite situation is also observed (e.g. Williams et al. 2014). Magnetic field measurements 
in unpolarised spectroscopy also reveal the existence of rapidly rotating stars with average 
surface fields well below the saturation value, taken as a hint of the fading of the rotation-dominated 
dynamo (Reiners \& Basri 2010). On their side, spectropolarimetric observations show the co-existence 
of two groups of stars with radically different types of large-scale magnetic fields within a narrow 
range of stellar parameters (Morin et al. 2010), which have been tentatively attributed to a 
bistability of the stellar dynamo (e.g. Gastine et al. 2013). It is however still a matter of debate 
to which extent these different observations are connected together, or for instance if some of 
them can be related to the physical conditions in the atmospheres of ultracool objects 
(e.g. Mohanty et al. 2002) or to the emission mechanisms (e.g. Hallinan et al. 2008).

During the past decade, large strides have been made in characterising and understanding the 
magnetic fields and activity of M dwarfs. However, our knowledge of these stars still remains 
very partial and several important puzzles are to be addressed. During the next few 
years, spectrographs and spectropolarimeters operating in the near-infrared will study extensively 
large samples of M dwarfs with the aim to discover rocky planets orbiting them (Delfosse et al. 2013; Oliva et al. 2014; Quirrenbach et al. 2014).
The synergy between these programs and studies of 
magnetic activity of low-mass stars -- with the need to model stellar activity both for planet 
detection and characterisation (e.g. H{\'e}brard et al. 2016; Donati et al. 2015; Ribas et al. 2016) -- are expected to 
contribute decisively to the next important advances in the field.

\subsection{Beyond the saturation: detecting the strongest magnetic fields in M dwarf stars}
\label{shulyak}

\begin{figure*}
\centerline{
 \includegraphics[width=0.75\hsize]{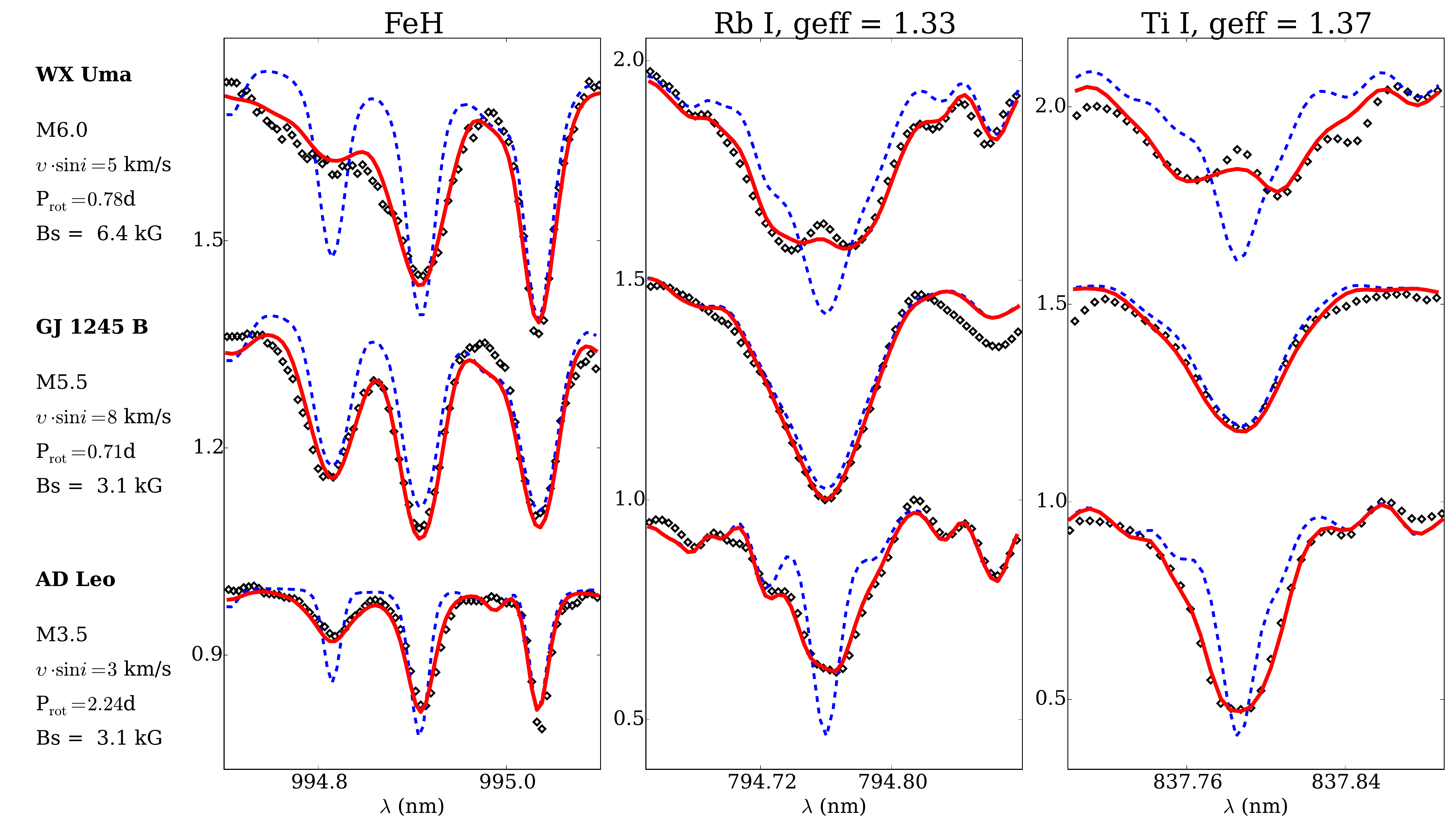}
}
\caption{
 \label{fig:line-fit}
  Detection of very strong magnetic field in WX~UMa.
  We show example fits to magnetic sensitive spectral lines in WX~UMa and a few other
  magnetic M dwarfs.
  Black dots~--~observed spectra; blue line~--~predicted zero field spectra;
  red line~--~best fit spectrum with the magnetic field. 
  The text on the left side of first column lists for each stars its name,
  spectral class, derived projected rotational velocity ($v\sin i$, where $i$ is the inclination angle
  between stellar rotation axis and the line-of-sight), rotational period,
  and the average surface magnetic field. Rotation periods are taken from 
  ZDI studies (Morin et al. 2008, 2010).
  In the case of WX~UMa one can see a clear Zeeman splitting in Rb~I and Ti~I lines, as well as
  a characteristic magnetic sensitive FeH feature at $994.8$~nm whose appearance
  corresponds to a very strong magnetic field of about $6.4$~kG.
}
\end{figure*}

Since the first detection of strong magnetic fields in M dwarfs in mid-80s (Saar \& Linsky 1985), several tens of individual measurements were obtained with different instruments 
and wavelength regions (Johns-Krull \& Valenti 1996, 2000; Reiners \& Basri 2007, 2010; Reiners et al. 2009; Kochukhov et al. 2009; Shulyak et al. 2014).
These measurements showed that the strength 
of the maximum possible surface magnetic field 
reaches values around 3--4~kG in the coolest M dwarf stars.
Fields stronger than this have not been detected in any low-mass star,
which was viewed as an evidence for the magnetic field saturation (Reiners et al. 2009),
similar to the saturation of stellar activity in terms of, e.g., 
X-ray fluxes which occurs for stars with rotational periods shorter 
than a few days (see, e.g., Shulyak et al. 2014).
Note, however, that all available measurements show 
large scatter between $2$~kG and $4$~kG, the latter being an upper
limit of the field that could be measured at that time with available 
techniques (Reiners \& Basri 2007).

Thanks to development of new analysis methods and techniques,
it is now possible to look at magnetic properties of stars in more detail.
In order to better understand the magnetism in M dwarfs
we used data collected over several years with the twin spectropolarimeters ESPaDOnS and NARVAL
mounted at the $3.6$~m Canada-France-Hawaii Telescope (CFHT) and the $2$~m Telescope Bernard Lyot (TBL) 
at the Pic du Midi (France), respectively (Petit et al. 2014). 
The ultimate aim of our work was to measure total magnetic flux from Stokes~$I$ spectra
by utilising up-to-date radiative transfer modelling and different spectroscopic diagnostics.

Our analysis resulted in the detection of the strongest $\langle B \rangle \approx6.4$~kG average M dwarf magnetic field
known to date in the star WX~UMa.
Figure~\ref{fig:line-fit} demonstrates example fits to magnetic sensitive spectral lines
in this and in several other magnetic M dwarfs which we have chosen for comparison purpose.
Thus WX~UMa is the first M dwarf star for which the Zeeman splitting in single atomic lines
is clearly observed at short wavelengths.

\begin{figure}
\centerline{
 \includegraphics[width=\hsize]{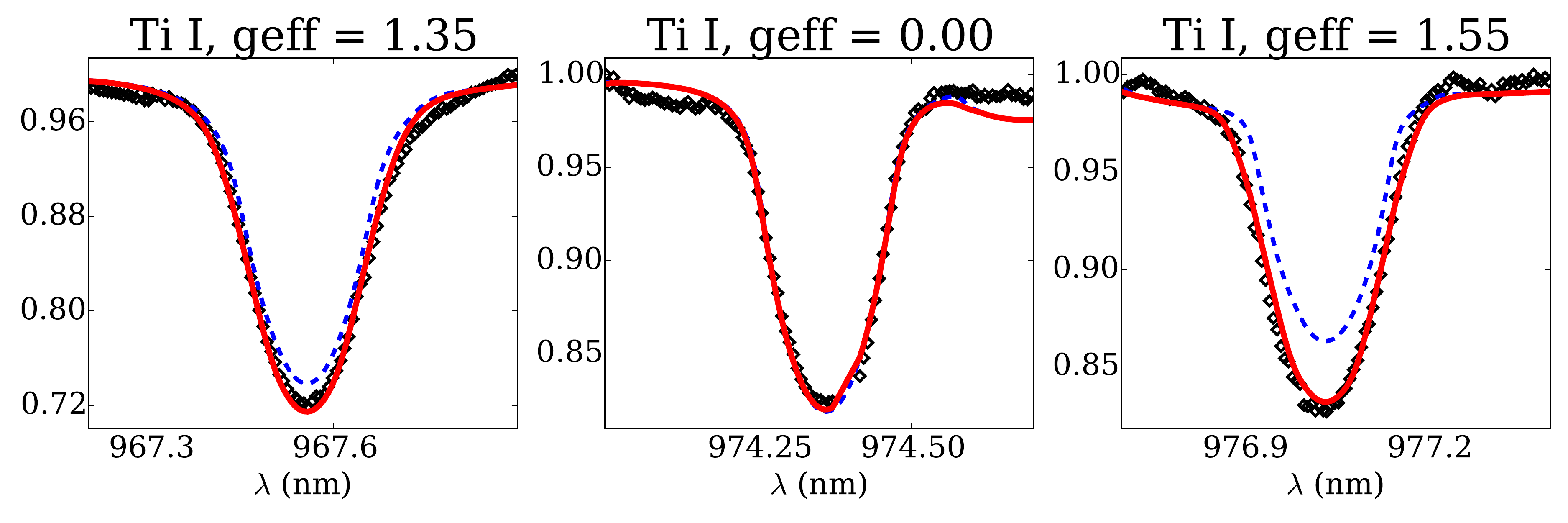}
}
\caption{
 \label{fig:maginten}
Example fit to \ion{Ti}{i} lines in the spectra of fast rotating M dwarf V374~Peg
with $v\sin i\approx35$~km\,s$^{-1}$.
Black diamonds~--~observations; full red line~--~best fit model with $\langle B \rangle=5.5$~kG;
dashed blue line~--~zero field model. The effective Land\'e g-factor ($g_{\rm eff}$) is
listed for each line in the title of corresponding subplot.
}
\end{figure}

We have also measured magnetic fields above $5$~kG in three other stars GJ~51, EQ~Peg~B, and V374~Peg.
However, these stars have rather large $v\sin i$ values 
which makes it impossible to see Zeeman
splitting in individual lines.
Thus, to measure magnetic fields in these stars we relied on the
effect of magnetic intensification of spectral lines (Landi Degl'Innocenti \& Landolfi 2004)
which predicts that the depth of a magnetic sensitive line broadened by rotation
will be increased depending on its Zeeman pattern. 
This technique is sensitive only to fields that are strong enough to produce
observable changes in the equivalent widths of spectral lines. 
We found that \ion{Ti}{i} lines in $\lambda\lambda$960--980~nm region are
superior diagnostic of strong fields in fast rotating stars
and we successfully used magnetic intensification of these lines in our analysis.
The reason why these lines are normally completely ignored in spectroscopic
studies is because they are heavily contaminated by telluric
absorption from the Earth atmosphere. Considering the importance of these
lines for our goals we made an effort of removing telluric absorption from all stellar spectra 
of all stars by applying the \textsc{MolecFit}
software package (Smette et al. 2015; Kausch et al. 2015).
As an example, our Figure~\ref{fig:maginten} demonstrates the model fit
to selected \ion{Ti}{i} lines in M dwarfs star V374~Peg which has $v\sin i\approx35$~km\,s$^{-1}$.

Thus, we reported for the first time
a detection of the magnetic fields in M dwarfs beyond the ``classical'' saturation limit
which was previously believed to be $\approx4$~kG.
Our finding provides an important constraint for the stellar dynamo theory.


\section{Conclusions}
\label{conclusions}

Results of the stellar magnetic field studies presented here emphasize the value of obtaining complementary constraints using different methods. Specifically, it is necessary to combine the Zeeman broadening analysis of intensity spectra which measures the total magnetic flux with the polarisation analysis of the large-scale field geometry. We have seen that it is not always straightforward to reconcile the two diagnostic methods, especially for the low-mass, fully convective dwarfs. The implications of this disagreement, for example the likely co-existence of the small and large-scale magnetic field structures at stellar surfaces, need to be taken into account by observational studies and addressed by theoretical models. 

At the same time, the comparison of the interferometric spot imaging with indirect Doppler spot mapping has provided encouraging results for a few cool stars accessible to interferometry. In particular, the question of the reality of a cool polar spot, debated for many years, appears to be settled. A combination of interferometric information with the traditional spectroscopic Doppler mapping of cool spots is a promising future direction for lifting the degeneracies of the latter technique and deriving some of the more uncertain star spot parameters (e.g. temperature contrasts).

We have witnessed a considerable progress in polarimetric field detections and analysis of main-sequence solar-like stars. Magnetic fields have been detected in hundreds of stars and mapped using ZDI in dozens of objects. A few stars show evidence of cyclic evolution of the field strength and topology, which is many cases not compatible with the activity cycles established from indirect proxy observations. In this context, recently reported discoveries of stars demonstrating a coherent, solar-like evolution of different proxies and direct magnetic fields indicators are particularly noteworthy.

Magnetic studies of young stars, especially open cluster members, are providing novel constraints on magnetic fields during the early stages of stellar evolution. These constraints are critical for understanding the shedding of angular momentum and activity decline as stars evolve towards the main sequence. The key science with the upcoming PEPSI polarimeter at the Large Binocular Telescope (LBT) will focus on solar-like  cluster stars. 

The magnetism of low-mass stars is arguably the most mysterious and difficult topic. The fully convective stars can't operate a tachocline dynamo and yet they exhibit a qualitatively similar rotation-activity relation as found in more massive solar-like stars. Moreover, the dynamo process operating in M dwarfs produces moderately strong, globally organised magnetic fields which yield easily detectable polarisation signatures. On the other hand, much stronger mean fields are evident from the Zeeman broadening analyses. The two types of magnetic measurements cannot be currently reconciled; considerable theoretical and modelling efforts are needed to resolve this puzzle.

The advancement of astronomical instrumentation has been the driving force behind most of the recent progress in understanding cool star magnetism. Forthcoming commissioning
of the high-resolution spectropolarimeter at the equivalent 11.8m diameter LBT (PEPSI) and of the first high-resolution night-time near-infrared spectropolarimeters (Spirou, CRIRES+) will likely answer some of the currently open questions and also lead to new discoveries.

\acknowledgements
C.P. Folsom was supported by the grant ANR 2011 Blanc SIMI5-6 020 01 ``Toupies: Towards understanding the spin evolution of stars'' ({http://ipag.osug.fr/Anr\_Toupies/}).
He also thanks the IDEX initiative at Universit\'e F\'ed\'erale Toulouse Midi-Pyr\'en\'ees (UFTMiP) for funding through the STEPS collaboration program between IRAP/OMP and ESO.
R. Fares acknowledges partial support from STFC consolidated grant number ST/J001651/1.
D. Shulyak acknowledges support from DFG project CRC~963 ``Astrophysical Flow Instabilities and Turbulence''.
O. Kochukhov acknowledges support by the Swedish Research Council and the Swedish National Space Board.
H. Korhonen thanks the Fonden Dr. N.P. Wieth-Knudsens Observatorium for the travel grant that made it possible for her to attend Cool Stars 19.  J.D. Monnier and R.M. Roettenbacher acknowledge support for the interferometric imaging work by the National Science Foundation (NSF) grant AST-1108963.

\end{document}